\documentclass[journal,twoside,web]{ieeecolor}
\usepackage{generic}
\usepackage{cite}
\usepackage{amsmath,amssymb,amsfonts}
\usepackage{algorithmic}
\usepackage{graphicx}
\usepackage{algorithm,algorithmic}
\usepackage{hyperref}
\hypersetup{hidelinks=true}
\usepackage{textcomp}
\usepackage[table]{xcolor}
\usepackage{float}
\usepackage{makecell}
\usepackage[T1]{fontenc}
\usepackage{pgfplots}

\usepackage{float}   
\usepackage{stfloats}   

\usetikzlibrary{patterns}

\def\BibTeX{{\rm B\kern-.05em{\sc i\kern-.025em b}\kern-.08em
    T\kern-.1667em\lower.7ex\hbox{E}\kern-.125emX}}
\begin{document}
\title{Sex Differences in the Detection of Parkinson's Disease from Speech}
\author{Hiba Akhaddar, Peter Plantinga, Mirco Ravanelli,  Tristan Glatard
\thanks{This paragraph of the first footnote will contain the date on 
which you submitted your paper for review.}
\thanks{Hiba Akhaddar was with Concordia University, Montreal, QC H3G 1M8, Canada, and Mila -- Quebec AI Institute, Montreal, QC H2S 3H1, Canada. She is now with the Krembil Centre for Neuroinformatics, Centre for Addiction and Mental Health (CAMH), Toronto, ON M5T 1R8, Canada (e-mail: h.akhaddar@gmail.com).}
\thanks{Peter Plantinga is with the Department of Neurology and Neurosurgery and the Centre for Research on Brain, Language and Music, McGill University, Montreal, QC H3A 2B4, Canada, and also with Mila -- Quebec AI Institute, Montreal, QC H2S 3H1, Canada.}
\thanks{Mirco Ravanelli is with the Department of Computer Science and Software Engineering, Concordia University, Montreal, QC H3G 1M8, Canada, and also with Mila -- Quebec AI Institute, Montreal, QC H2S 3H1, Canada.}
\thanks{Tristan Glatard was with the Department of Computer Science and Software Engineering, Concordia University, Montreal, QC H3G 1M8, Canada. He is now with the Krembil Centre for Neuroinformatics, Centre for Addiction and Mental Health (CAMH), Toronto, ON M5T 1R8, Canada.}}

\maketitle
\begin{abstract}
  Parkinson's disease (PD) is the second most common neurodegenerative disease in the world, and 70\% to 75\% of PD patients suffer from speech disorders at some stage. Since PD affects men more frequently than women, previous publications tend to rely on unbalanced datasets that potentially introduce sex biases into trained models. 
This study investigates sex differences in PD detection from speech using deep learning and traditional machine learning approaches. We evaluated speech-based PD detection using embeddings from three pre-trained speech models, combined with handcrafted acoustic features and different classifiers. Sex-specific experiments were conducted on balanced subsets of four sustained phonation datasets, controlling for sex, disease status, and, where possible, age. All recordings were obtained during a sustained phonation task. Results show that female speakers were more accurately classified than male speakers in the larger dataset. Sex-specific results show that when the models are trained on one sex, they fail to generalize to the other one. Our results highlight the importance of including sex as a variable in the development of PD detection systems.
\end{abstract}


\section{Introduction}
\label{sec:introduction}
\IEEEPARstart{P}{arkinson's} disease (PD) is a neurodegenerative disease that affects more than 10 million people in the world, making it the second most common neurodegenerative pathology~\cite{dorsey2018parkinson}. PD is associated with a deficiency in dopamine production resulting from the degeneration of neurons in the midbrain region. Low dopamine levels cause symptoms like involuntary motion, muscle rigidity, and alterations in speech and handwriting~\cite{jankovic2008parkinson}. 

Although no curative treatments have yet been developed for the pathology, providing an early and accurate diagnosis is crucial \cite{poewe2017parkinson} for better disease monitoring, optimizing patient care, and advancing research efforts. 

The UK Parkinson's Disease Society Brain bank and the Movement Disorders Society’s diagnostic criteria showed that the clinical diagnosis of PD can be done based on bradykinesia and one of the following features which are rigidity, rest tremor, and postural instability \cite{iyer2023machine}. However, voice and speech alterations can appear as early as five years before the onset of major motor symptoms \cite{reddy2024automaticy}, while diagnosis is often delayed until these motor symptoms become evident \cite{postuma2015mds}. This gap highlights the potential of speech-based approaches for early and accessible PD detection. 

Speech impairment is one of the most common motor manifestations of PD, affecting 70 to 90 \% of patients \cite{logemann1978frequency} and frequently presenting as hypokinetic dysarthria, characterized by reduced vocal loudness, monopitch, imprecise articulation, and altered prosody \cite{skodda2010speech}. These abnormalities result from the neurodegenerative processes associated with PD, which impair the motor control of the respiratory, phonatory, and articulatory systems involved in speech production \cite{ho2012speech}. Consequently, patients may exhibit progressive deterioration in speech, including increased hesitancy, dysfluency, and, in advanced stages, near-mutism, leading to reduced speech intelligibility and impaired communication~\cite{bib4}.

Recent works investigated the use of Deep Learning models for PD detection from speech because it can capture complex information directly from raw data \cite{Simone2025InterpretableEarly}. 
These models are more affordable compared to models that use other modalities and they can be deployed remotely through smartphones and telemedicine platforms. These advantages are important given the growing rate of PD which requires scalable diagnostic tools.

However, PD does not manifest uniformly across patients. Although based on epidemiological data, PD is 1.5 times more prevalent in men \cite{haaxma2007gender}, the disease tends to progress at a faster rate in females \cite{bib2}. These sex-related differences suggest that PD manifests heterogeneously across patients and highlight the importance of developing diagnostic approaches that are robust to demographic variability.

Studying sex differences is crucial since natural speech patterns differ between healthy men and women, including fundamental frequency, which typically averages around 120~Hz for males and 220~Hz for females \cite{hunter2010variations}, and formant frequencies \cite{gelfer2016speaking}, which are the resonant frequencies of the vocal tract. These variations in acoustic properties could influence how PD may manifest differently across sexes. In \cite{ahn2023sex}, the authors showed that speech characteristics associated with dysphagia differ by sex. Hypophonia was more prevalent among men, while imprecise articulation was more common among women. However, many existing studies overlook these differences, often by relying on unbalanced datasets skewed towards male patients, which could lead to biased models and less generalizability \cite{wang2021algorithmic}.

Despite the growing success of speech-based PD detection systems, few studies have systematically investigated how differences related to sex affect model performance across datasets and learning architectures. It remains unclear, though, whether models trained on one sex generalize equally to the other sex or whether demographic composition of the training data introduces biases that are sex specific. It is also uncertain whether performance differences observed on large scale speech datasets persist when demographic composition and dataset size are controlled.

The goal of this paper is to investigate differences related to sex in Parkinson's disease detection from speech while controlling for potential confounding factors such as age and dataset size.

The distinctive contributions of our work are summarized as
follows:

\begin{enumerate}
    \item We analyzed sex differences in Parkinson's disease (PD) detection from speech across multiple datasets and model architectures, comparing self-supervised representations with traditional acoustic features combined with Time-Delay Neural Network and Random Forest classifiers.
    \item We studied the effects of dataset composition from those of model generalization through controlled experiments on subsets of datasets that varied demographic matching, training set size, and the sex of training and test speakers.
    \item Based on our findings, we provided recommendations for dataset construction and model evaluation to promote fairness in future clinical speech AI systems.
\end{enumerate}

\section{Related Work}

\subsection{Detecting Parkinson's Disease from Speech using machine and deep learning}

Several studies have demonstrated that Parkinson's disease can be detected from speech using both traditional Machine Learning (ML) and Deep Learning (DL) methods. 
Earlier approaches relied on handcrafted acoustic features combined with classifiers, whereas recent studies increasingly include self-supervised speech representation models. 
These methods have been successfully applied to sustained phonation, diadochokinesis (DDK), reading, and spontaneous speech tasks, consistently reporting high detection performance across multiple datasets \cite{Almeida2019Detecting,bib3,LaQuatra2025Bilingual,GimenoGomez2024Unveiling,Simone2025InterpretableEarly,Chronowski2022Parkinson}.

Recent work has focused on improving robustness, interpretability, and generalization. For example, La Quatra et al. showed that combining speech enhancement with self-supervised models improved performance on real-world recordings, while another study proposed a dual-head architecture capable of generalizing across Spanish and Polish datasets \cite{bib3,LaQuatra2025Bilingual}. Other papers incorporated attention mechanisms to improve interpretability or investigated architectures that were based on transformers for PD detection \cite{GimenoGomez2024Unveiling,Simone2025InterpretableEarly}. Mancini et al. systematically evaluated post-hoc explainability methods, including perturbation and saliency maps, on a speech-based PD detector \cite{Mancini2025Investigating}.

Collectively, these studies demonstrate that modern self-supervised representations outperform many traditional feature extraction techniques and have become the dominant approach for diagnosing PD from speech.

Despite these advances, current studies primarily focus on maximizing classification accuracy and comparing model architectures. Their evaluations are performed on a single dataset or emphasize cross-dataset and cross-lingual generalization. The effect of biological sex on speech-based PD detection has not been systematically investigated, motivating the analysis presented in the next subsection.

\subsection{Sex differences in the detection of Parkinson's Disease from Speech}

Several studies reported conflicting results regarding the impact of sex on the accuracy of PD detection from speech. While some work report higher performance for male speakers, others observe better performance for female speakers.

Table \ref{tab:sex_studies} summarizes the main studies that have investigated the influence of biological sex on speech-based Parkinson's disease detection. Although relatively few works explicitly analyze sex differences, their findings consistently indicate that sex affects both speech characteristics and model performance.

\begin{table*}[h!]
\centering
\caption{Summary of studies investigating sex differences in PD detection from speech} \label{tab:sex_studies}
\begin{tabular}{|p{1.5cm}|p{2.5cm}|p{3cm}|p{4cm}|p{4cm}|}
\hline
\textbf{Author's name [ref.]} & \textbf{Dataset subjects} & \textbf{Methodology} & \textbf{Study Result (best performance)} & \textbf{Justification of Sex Differences} \\
\hline
Houle et al. \cite{houle2023sex} & 68 speakers, sex-balanced and disease-balanced & Logistic regression on acoustic features & Women. Women had more distinguishable acoustic patterns. Sex affected smoothed cepstral peak prominence, net syllables/sec, percent pause ratio, articulatory-acoustic vowel space & Acoustic measures differed by sex; release burst precision differed: females produced multiple bursts more often, males produced no burst more often. \\
\hline
Rusz et al. \cite{bib2} & 100 speakers, 60 men and 40 women PD patients, matched HC & Logistic regression on phonation, articulation, and prosody features & Women. Higher speech abnormality prevalence for women (ROC 65\% vs 56\%) and higher classification accuracy (AUC 0.93 vs 0.86) & Voice quality and articulation features favor female detection; loudness variability favors male detection. \\
\hline
van Gelderen et al. (survey) \cite{vangelderen2024deep} & 33 DL studies reviewed & Systematic review of deep learning studies on PD detection from speech & Mixed. Some studies report higher performance for men, others for women & Bias may arise from dataset composition, sex imbalance, or feature distributions. \\
\hline
Jeancolas et al. \cite{Jeancolas2021XVectors} & 221 speakers, sex-balanced, more PD than HC & MFCC-GMM baseline; Cosine similarity, LDA, PLDA on x-vector embeddings  & Men. Better performance with an Equal Error Rate (EER) of 22\% vs. 30–39\% for women. & Men exhibit lower inter-speaker acoustic variability, making their speech patterns easier for models to classify. \\
\hline
Khaskhoussy et al. \cite{khaskhoussy2021deep} & 4 datasets & SVM and LSTM classifiers on acoustic, prosodic, phonetic features & Men. LSTM led to the highest performance for men F1 81–99\% & Bias could arise because datasets contained more men than women. \\
\hline
Adnan et al. \cite{Adnan2025Fusion} & 1306 speakers, 614 men and 693 women, 392 PD and 914 HC & Neural classifier on combined Wav2Vec2.0, WavLM, ImageBind embeddings & Women. Higher detection accuracy (86.9\% vs. 84.3\% for men). & Female PD patients have more consistent and pronounced acoustic deviations, aiding detection. \\
\hline
\end{tabular}
\end{table*}

Several studies have shown that speech impairments related to PD manifest differently in men and women. In \cite{houle2023sex}, Houle et al. reported that women exhibited more distinguishable acoustic patterns, with sex influencing measures such as cepstral peak prominence, speech rate, pause ratio, and articulatory-acoustic vowel space. Similarly, in \cite{bib2},  Rusz et al. found a higher prevalence of speech abnormalities in women and observed that different speech dimensions were affected depending on sex. These findings indicate that sex influences the acoustic manifestation of PD rather than simply acting as a demographic variable.
  
Despite evidence that PD affects male and female speech differently, the impact of sex on automatic PD detection remains unclear. Some studies reported higher performance for male speakers, such as the work by Jeancolas et al.\cite{Jeancolas2021XVectors} and Khaskhoussy et al. \cite{khaskhoussy2021deep}, whereas others achieved better performance for female speakers, such as the paper by Adnan et al. \cite{Adnan2025Fusion}. Proposed explanations include differences in inter-speaker acoustic variability, dataset imbalance, and sex-dependent acoustic deviations, but no consensus has emerged regarding the dominant contributing factors. Therefore, the direction of the observed sex differences appears to depend on both the dataset and the modeling approach.

As highlighted by the survey of van Gelderen et al. \cite{vangelderen2024deep}, sex-specific analyses remain uncommon in the speech-based PD literature, and most studies report overall performance without systematically evaluating fairness across demographic groups. Consequently, it is still unclear whether the observed sex differences arise from biological differences in speech production, dataset composition, model architecture, or interactions among these factors.

To address this gap, our work systematically evaluates the effect of sex across multiple datasets, feature representations, and model architectures while controlling for demographic confounding factors through balanced experimental protocols.

\section{Methods}

We evaluated multiple speech representation approaches across three datasets. We
compared self-supervised representations,
spectral features, and handcrafted acoustic features using
deep learning and traditional classifiers. Experiments were conducted on three
datasets: mPower~\cite{bot2016mpower}, PC-GITA~\cite{orozco2014new} and QPN (Quebec Parkinson Network) \cite{bib6}. We also included controlled subsets of mPower designed to study the
effects of demographic matching and dataset size. Cross-sex training and testing
protocols were further used to assess model generalization across sexes.

\subsection{Models}

\begin{figure*}
\centering
\includegraphics[width=\textwidth]{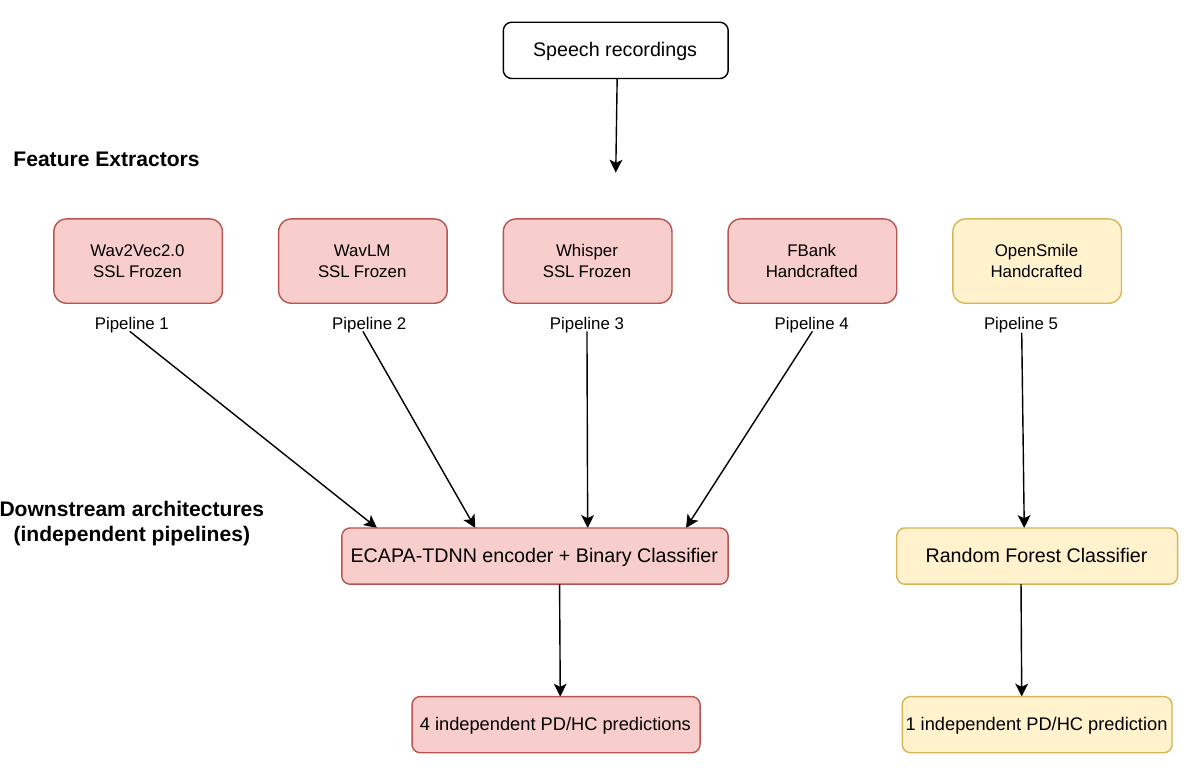}

\caption{Comparative feature extraction approaches for Parkinson's disease detection}
\label{fig:1}
\end{figure*}

Figure \ref{fig:1} summarizes the four speech processing pipelines evaluated in
this work. To study sex differences, we compared self-supervised and
conventional feature representations while controlling for demographic factors
and using the same downstream classifier whenever applicable. Three pipelines
use frozen self-supervised speech encoders namely Wav2Vec2.0
\cite{baevski2020wav2vec}, WavLM \cite{chen2022wavlm}, and Whisper
\cite{radford2023robust}, whose embeddings are classified using an ECAPA-TDNN
network. Additionally, 80-dimensional FBank features are evaluated with the same
ECAPA-TDNN architecture to compare learned and conventional spectral
representations. As a traditional baseline, handcrafted ComParE2016 acoustic
features are classified using a Random Forest.
This comparison allows assessment of whether large-scale pretrained models offer
representational benefits over handcrafted acoustic descriptors.

The pretrained self-supervised learning (SSL) models were chosen because they capture nuanced acoustic and phonetic patterns that are potentially relevant to classify between PD and HC. These embeddings encapsulate speech dynamics such as articulation rate, voice tremor, and phonatory instability, which are subtle and could be indicative of PD \cite{tsanas2012}.

Wav2Vec2.0 \cite{baevski2020wav2vec} learns speech representations by masking portions of the input and using a contrastive loss to identify the correct features. It was pretrained on approximately 60,000 hours of unlabeled speech from the LibriSpeech \cite{panayotov2015librispeech}, Libri-Light \cite{kahn2020libri}, and CommonVoice \cite{ardila2020common} datasets. The model uses a convolutional encoder that processes raw audio waveforms at a 16 kHz sampling rate, producing latent speech representations at a 50 Hz frame rate. The context network, implemented as a Transformer architecture \cite{vaswani2017attention}, generates contextualized representations that capture temporal dependencies across the utterance.

WavLM \cite{chen2022wavlm} extends this approach by adding gated relative position bias and utterance mixing to capture phonetic, acoustic, and prosodic information. The model was pretrained on 94,000 hours of public audio data using a combination of masked prediction and denoising objectives. These enhancements allow better modeling of speaker characteristics and acoustic variations, making it suitable for tasks involving pathological speech \cite{vasquez2018towards}. We used WavLM-Large.

Whisper \cite{radford2023robust} was trained on approximately 680,000 hours of multilingual and multitask supervised data for automatic speech recognition. The encoder produces rich cross-lingual embeddings that capture acoustic representations across diverse recording conditions and speaker populations. We specifically used the Whisper-large-v2 models, which contain approximately 1.55 billion parameters and are based on a Transformer-based encoder-decoder architecture.

In contrast, FBank captures the spectral content of the speech waveform through short-time Fourier transform followed by mel-filters applied on the frequency axis.

The ComParE2016 feature set, that includes 6373 handcrafted acoustic features, was extracted using OpenSMILE toolkit \cite{eyben2010opensmile}. The ComParE2016 features which include foundational frequency parameters such as F0, jitter, and shimmer, which are different between men and women in healthy speech influence the sex specific pattern learning.
Unlike FBank features, which preserve frame-level temporal information and can therefore be processed by ECAPA-TDNN, ComParE2016 consists of utterance-level statistical functionals that do not retain temporal structure. Consequently, ComParE2016 features were classified using a Random Forest, following the standard evaluation protocol for this feature set.

All pretrained SSL encoders were used in a frozen configuration to preserve the representational quality of pretrained speech embeddings, avoid task-specific adaptation, and investigate the transferability of pretrained features to pathological speech data. This design isolates the contribution of the learned speech representations themselves. We ensure that differences in downstream performance arise solely from representation quality, not from fine-tuning differences. This allows us to get a better comparison across models. 
Moreover, freezing the encoder weights allows us to significantly reduce the computational requirements during training and mitigate the risk of overfitting which is essential considering the limited availability of labeled pathological speech data.

For the SSL embeddings and FBank features, the extracted representations are fed to an ECAPA-TDNN classifier. ECAPA-TDNN  network \cite{desplanques2020ecapa} was originally designed for speaker recognition, but has been used for many other speech processing tasks like speaker diarization \cite{Dawalatabad2021ECAPATDNN}, language classification \cite{Chaitra2023ECAPA_TDNN_LID}, and text-to-speech \cite{Xue2022ECAPA_TDNN_TTS}. ECAPA-TDNN was followed by a fully connected binary classification layer to predict disease status. 

 The embeddings from each feature extractor were fed in turn to ECAPA-TDNN network, configured with an input size of 1024 for Wav2Vec2.0, WavLM, and FBank, and 1280 for Whisper to accommodate its higher-dimensional representations, 5 convolutional blocks with channel configuration of [1024, 1024, 1024, 1024, 3072], kernel sizes of [5, 3, 3, 3, 1], and dilations of [1, 2, 3, 4, 1]. The network employed 128 attention channels and output a 192-dimensional utterance-level embedding. These architectural parameters were selected based on prior work on robust speaker representation learning  \cite{Dawalatabad2021ECAPATDNN} and our experimental results.

ECAPA-TDNN transforms variable-length audio recordings into fixed-dimensional vectors for classification. The network applies one-dimensional convolutional layers to extract temporal patterns from the audio, using residual connections \cite{he2016deep} to preserve information across layers and dilations to capture long-range temporal dependencies. The attentive pooling mechanism summarizes the sequence into a single representation, giving greater weight to frames that contain the most informative speech characteristics for PD detection. The resulting fixed-dimensional vectors were passed to a fully connected classifier with log-softmax activation that converts the logits to log-probabilities for binary prediction. We trained using the negative log-likelihood loss that maximizes the likelihood of the correct class.

Unlike the SSL and FBank pipelines, ComParE2016 features were classified using Random Forest (RF) as a baseline to benchmark against deep learning architectures.

\subsection{Datasets}

We ran experiments on three datasets namely mPower~\cite{bot2016mpower}, PC-GITA~\cite{orozco2014new} and QPN (Quebec Parkinson Network) \cite{bib6}. Their demographic and clinical characteristics are summarized in Table~\ref{tab:demographics}. We obtained ethical approval through Concordia University on file number 30021546 and by getting the Tri-Council Policy Statement: Ethical Conduct for Research 
Involving Humans (TCPS 2: CORE 2022) certificate in order to access them. 
\begin{table*}[t]
\centering
\caption{Demographic and clinical characteristics of the datasets used in this study. Values represent mean $\pm$ standard deviation for continuous variables. Age is expressed in years.}
\label{tab:demographics}
\renewcommand{\arraystretch}{1.2}
\resizebox{\textwidth}{!}{%
\begin{tabular}{|p{3.5cm}|p{2.2cm}|p{2.2cm}|p{2.2cm}|p{2.2cm}|p{2.2cm}|}
\hline
& 
\textbf{mPower large} & 
\textbf{mPower matched} & 
\textbf{mPower small} & 
\textbf{PC-GITA} &
\textbf{QPN}\\
\hline
$n$ samples                & 690 & 452 & 100 & 100 & 100\\
Male/Female (\%)           & 50/50 & 50/50 & 50/50 & 50/50 & 50/50\\
PD/HC (\%)           & 50/50 & 50/50 & 50/50 & 50/50 & 50/50\\
Age (male, mean $\pm$ SD)   & 44.34 $\pm$ 8.63 & 59.53 $\pm$ 7.06 & 59.53 $\pm$ 7.06 & 61.08 $\pm$ 7.05 & 66.18 $\pm$ 12.77\\
Age (female, mean $\pm$ SD) & 51.38 $\pm$ 8.26 & 59.55 $\pm$ 6.86 & 59.55 $\pm$ 6.86 & 61.92 $\pm$ 11.35 & 67.86 $\pm$ 6.26 \\
UPDRS (male, mean $\pm$ SD) & --- & --- & --- & 37.76 $\pm$ 22.09 & 28.40 $\pm$ 12.26 \\
UPDRS (female, mean $\pm$ SD) & --- & --- & --- & 37.56 $\pm$ 14.03 & 28.64 $\pm$ 10.94 \\
\hline
\end{tabular}
}

\end{table*}

The mPower dataset \cite{bot2016mpower} is a large publicly available dataset collected by Sage Bionetworks in collaboration with the University of Rochester through the mPower mobile health application. Volunteers remotely recorded themselves doing the sustained phonation tasks by saying the vowel /a/ 'Aaah' for 10 seconds. The recordings were obtained using the built-in microphones of the participants' iPhones at a sampling rate of 44.1 kHz. Although the dataset includes over 65,000 participants who performed different tasks, the demographic distribution is not uniform across age, sex, and health status.

To investigate the influence of dataset size and demographic matching, from the mPower dataset, we extracted three subsets, which are presented in Table \ref{tab:demographics}. 
The first subset (mPower large) has the largest number of samples but lacks age matching between males and females, which reflects real-world conditions where demographic imbalances are common in health datasets \cite{coravos2019developing}. 

However, the second (mPower matched) is smaller in size but is age-matched, which constitutes a suitable set for direct comparison with the PC-GITA dataset. Age matching was performed by restricting participants to the 50 to 70 age range and ensuring the mean age difference was less than 1 year. The controlled demographics allow us to study the effect of model architecture and feature representation from potential age-related effects.

The third subset (mPower small), with 100 participants to match the sample size of PC-GITA, allows us to evaluate the model's performance under identical data constraints. The difference with mPower matched is only the dataset size; both sex and age are matched as well.

PC-GITA is a clinically validated corpus of Spanish speakers collected at the Clinica Noel Hospital in Medellín, Colombia. It includes several speech tasks, which are sustained phonation, reading, and spontaneous speech. Recordings were acquired in a recording room using a dynamic omnidirectional microphone at a sampling rate of 44.1 kHz. The dataset contains 100 participants, equally divided between males and females and between PD and HC subjects. In addition, PD and HC participants are age-matched, and disease severity was assessed using the Unified Parkinson's Disease Rating Scale (UPDRS-III).
In our work, we restricted sustained phonation to the vowel /a/ to be consistent with the mPower data and because it is widely studied in voice analysis as it reflects motor impairments in PD.

The Quebec Parkinson Network (QPN) dataset~\cite{bib6} consists of speech recordings collected from individuals with Parkinson's disease and healthy controls in Quebec, Canada. QPN is a set of speech recordings that includes 208 PD patients (133 men and 75 women) and 75 HC (25 men and 50 women). For consistency across datasets, we selected a balanced subset of 100 participants (50 PD and 50 HC, with equal numbers of males and females). The participants were recorded in a quiet room using a headset microphone near the mouth. We used the sustained phonation task and all the recordings have been resampled to 16 kHz prior to use.

\subsection{Model Training}

The deep learning models were implemented using PyTorch and SpeechBrain \cite{ravanelli2021speechbrain} and trained with the Adam optimizer at a learning rate of $1 \times 10^{-4}$ and a weight decay of $2 \times 10^{-5}$.
A cyclic learning rate scheduler with exponential decay was employed to improve training stability and reduce overfitting, with hyperparameters selected empirically. All models were trained for 30 epochs with a batch size of 16 and gradient accumulation of 2, resulting in an effective batch size of 32 while staying within GPU memory limits. The final test evaluation used the model checkpoint with the loweokst validation error. To ensure reproducibility, random seeds were fixed across experiments, and all pretrained models were obtained from the Hugging Face Transformers library.

We then reported the overall and sex-specific F1-scores, accuracy, precision, and recall on the test set.

Since WavLM-large achieved the strongest performance among SSL feature extractors on the mPower large dataset, we used it as the only feature extractor on the matched datasets. This experimental design allows us to determine whether the observed performance trends remain consistent when dataset size and demographic composition are controlled. The same evaluation metrics were reported for all the datasets and experiments. 

To evaluate whether our models performed above chance level, we conducted a permutation test. A null distribution was generated by randomly assigning class labels to the test samples and computing the performance metric for each random assignment. The observed model performance was then compared with this null distribution to estimate the statistical significance of the results.

To investigate whether sex-specific characteristics influence model generalization, we evaluated cross-sex performance by training separate models on male-only and female-only cohorts and testing them on both sexes. This analysis examines whether models trained on one sex generalize equally to the other sex and whether they exhibit differences in PD detection performance across sexes.

To study the effect of age, we used PC-GITA and a subset from mPower. We implemented a matched-pairs algorithm to obtain a subset in which age, sex, and disease status were matched to PC-GITA subjects. The two cohorts differ in context: PC-GITA was clinically validated and recorded in a laboratory setting, whereas mPower samples were collected from volunteers, reflecting a real-world setting. This allows us to compare the robustness of features across different recording conditions.

We used QPN as a replication cohort, in which the dataset is balanced for age and sex and disease severity is matched. The population is older, but the disease severity is lower than in PC-GITA.

\section{Results}

The results of the experiments are presented in Figures \ref{fig:feature_extract}, \ref{fig:datasets_compare}, and Table \ref{tab:sex_specific_combined}.
All reported metrics represent mean ± standard deviation across cross-validation folds and random seeds. Statistical significance was assessed using permutation tests \cite{ojala2010permutation} comparing model performance against a random baseline. Significance levels are denoted as hached bars if not significant in the figures and in the tables * p < 0.05, ** p < 0.01, *** p < 0.001, after using the Bonferroni correction.

\subsection{Sex differences in PD detection performance on the mPower large dataset}

\begin{figure*}
\centering
\includegraphics[width=\textwidth]{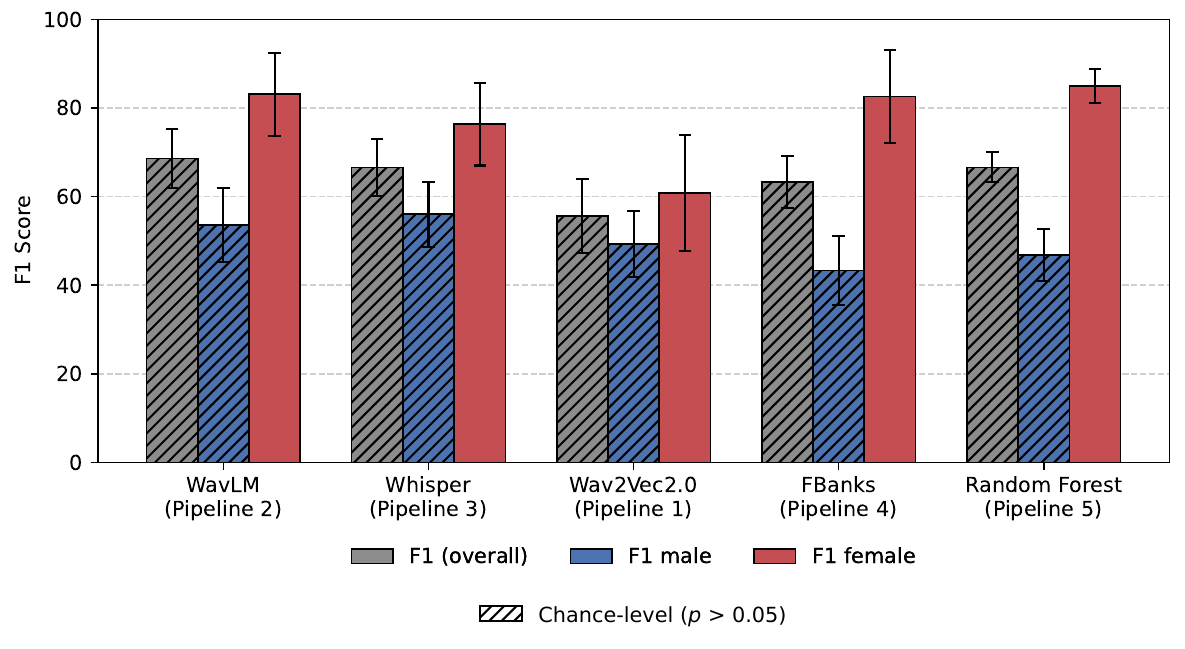}

\caption{Performance of 5 feature extractors on mPower large}
\label{fig:feature_extract}
\end{figure*}

To investigate whether sex affects PD detection performance, we evaluated all
feature extraction pipelines on the balanced mPower large dataset.
Figure~\ref{fig:feature_extract} summarizes the overall and sex-specific
F1-scores obtained across 10 folds and 5 random seeds. Accuracy, precision, and recall results are provided in
Appendix~\ref{app:results}, table ~\ref{tab:accuracy}.

The overall performance of the evaluated pipelines was close to chance level. However, sex-specific analysis revealed substantial differences between male and female participants.

Female-specific F1 scores were significantly above chance level (p<0.001), whereas none of the male-specific F1 scores reached statistical significance across all models. This indicates a pronounced sex-based disparity in PD detection. The best female-specific performance was achieved by Random Forest with ComParE2016 features (85.02\% $\pm$ 3.79), while the corresponding male F1 score remained substantially lower (46.80\% $\pm$ 5.83). Similar disparities were observed for FBanks (82.63\% $\pm$ 10.57 vs. 43.30\% $\pm$ 7.73) and the SSL models, including WavLM (83.12 \% ± 9.36) and Wav2Vec2.0 (60.79\% $\pm$ 13.05 vs. 49.30\% $\pm$ 7.39). These findings highlight the difficulty of learning robust representations for male participants in the mPower large dataset.

The observed sex disparity was consistent across both traditional ML approaches and SSL-based DL models. This shows that it is unlikely to be an artifact of a particular model architecture. Instead, it may reflect factors related to the dataset, including differences in sample size, recording conditions, and acoustic characteristics or biological differences between male and female participants. One potential explanation is the importance of fundamental frequency (F0), a known marker for PD detection \cite{sakar2013collection}. Female voices typically exhibit higher F0 values (165–255 Hz) than male voices (85–180 Hz) \cite{baken2000clinical}, which may partially explain the learning of discriminative decision boundaries for female recordings while limiting generalization to male speech.

Among the SSL models, WavLM achieved the highest overall performance, followed by Whisper and Wav2Vec2. This may be attributed to WavLM's pretraining, which combines masked speech prediction with denoising objectives and utterance mixing, enabling the learning of representations that capture phonetic information as well as speakers acoustic variations. These properties may be beneficial for pathological speech analysis, where subtle changes in voice quality, prosody, and articulation are important.

In contrast, Whisper achieved lower performance despite being pretrained on 680,000 hours of multilingual speech, suggesting that larger pretraining corpora alone do not necessarily translate into better performance for clinical speech tasks. Similarly, Wav2Vec2.0 showed the lowest performance among SSL models, possibly because its representations were learned primarily from clean read speech from LibriSpeech and may emphasize phonetic discrimination rather than pathological voice characteristics~\cite{skodda2011vowel,duffy2013motor}. In addition, our dataset included a sustained phonation task, which differs from the connected speech used in LibriSpeech. 

Random Forest with handcrafted ComParE2016 features outperformed all SSL models for female participants. This suggests that conventional acoustic features remain effective for pathological speech analysis and may capture disease-related characteristics that are not well represented in current SSL embeddings. A likely explanation is that existing SSL models are pretrained primarily on healthy speech from audiobooks, podcasts, and conversational recordings, creating a domain mismatch that limits transferability to pathological speech.

\subsection{Sex differences on different datasets and size effect}

Since the observed sex disparity was consistent across the evaluated pipelines, we next investigated whether it was influenced by dataset characteristics including dataset size, age distribution, and recording conditions. WavLM-Large was selected as the feature extractor for these experiments because it achieved the highest overall accuracy on the mPower large dataset (69.04\% $\pm$ 5.86; see Appendix~\ref{app:results}, table \ref{tab:datasets_dif}). For the smaller datasets, we used five folds and three random seeds. We present the results from using the WavLM pipeline on mPower matched and mPower small, in addition to the PC-GITA datasets.

\begin{figure*}
\centering
\includegraphics[width=\textwidth]{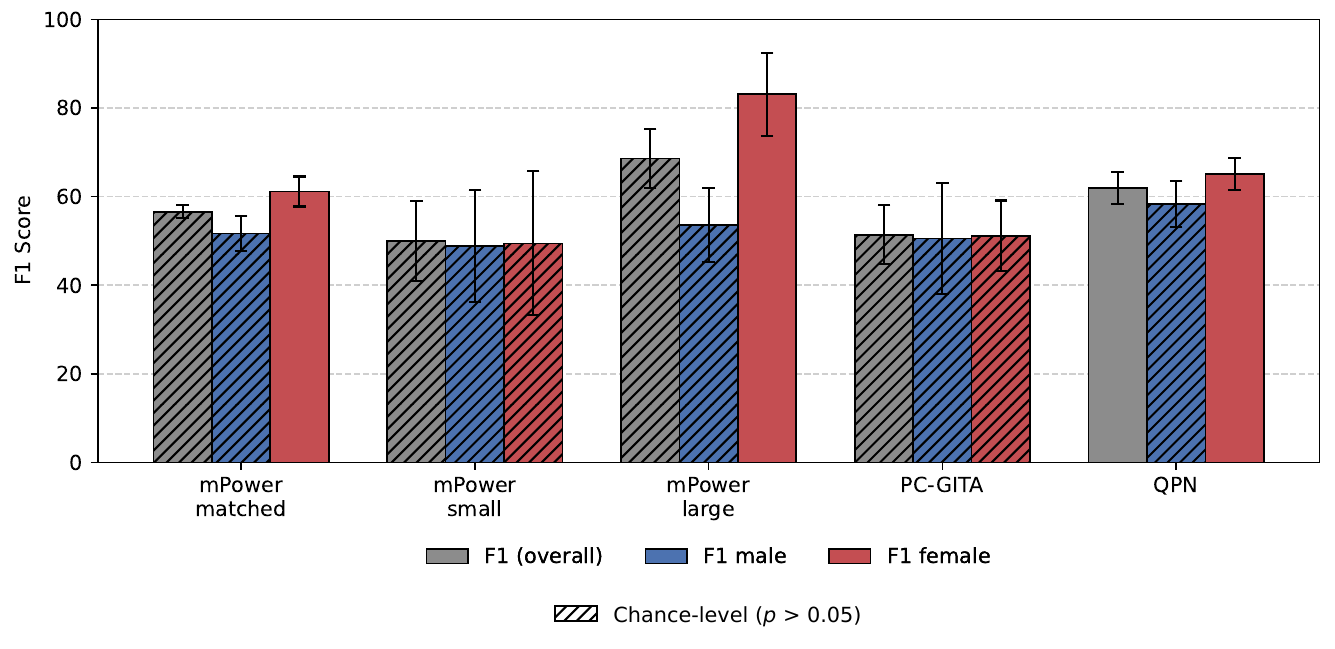}

\caption{Performance of WavLM across datasets}
\label{fig:datasets_compare}
\end{figure*}

Figure~\ref{fig:datasets_compare} summarizes the performance of the WavLM pipeline across the mPower, PC-GITA, and QPN datasets. The results reveal a consistent sex-specific pattern, which is that male-specific PD detection did not significantly exceed chance level in any dataset, whereas female-specific performance was significantly above chance in the larger datasets. 
On mPower large, the female-specific F1-score reached 83.12\% $\pm$ 9.36, while the male-specific F1-score was 53.65\% $\pm$ 8.39. Similarly, female-specific performance was significantly above chance on QPN (65.10\% $\pm$ 3.52), although the corresponding male-specific performance was not (58.30\% $\pm$ 5.23). The mPower matched subset also showed a significant female-specific performance advantage (61.16\% $\pm$ 3.39 versus 51.71\% $\pm$ 3.93 for males).

In contrast, neither the mPower small nor the PC-GITA dataset resulted in sex-specific performance significantly above chance. Both datasets contain only 100 samples and have controlled sex and age distributions. 
These results suggest that the observed female-specific performance advantage may depend on the amount of available training data. In fact, although female-specific detection was significant in the larger datasets, performance did not exceed chance in the smaller datasets. Across all datasets, male-specific performance remained at chance level, suggesting that the model's ability to detect PD from speech may be particularly limited for male participants under the conditions evaluated in this study.

Figure \ref{fig:datasets_compare} shows that sex related differences in performance were significantly above chance in the large datasets only, and the direction of the effect was always towards women. The results demonstrate that the sex difference could be dataset-dependent, with stronger performance for women in mPower datasets. The size of the dataset is also affecting the results. On mPower, the larger the dataset, the better the performance, and the larger the gap between men and women. The dataset demographic information is provided in Table \ref{tab:demographics}.

The performance on PC-GITA, where the recordings were collected in a laboratory, and mPower small is interesting as it is very similar to each other.
PC-GITA was collected in a controlled clinical setting, whereas mPower small consists of recordings collected remotely in a real-world environment. Moreover, PC-GITA has been successfully used for speech-based PD detection, indicating that its sample size is not inherently insufficient to distinguish PD from HC. Therefore, the lack of statistically significant performance in the present experiments cannot be attributed solely to the small sample size. Instead, it may reflect the combined effects of sample size, dataset composition, recording conditions, speech task, and population characteristics. These results suggest that the observed sex-specific performance differences are sensitive to the characteristics of the dataset and may not generalize uniformly across different clinical and real-world speech corpora.

\subsection{Performance of sex-specific training datasets}

To further investigate sex differences in the datasets, we conducted additional experiments by training on one sex and testing on both sexes. We evaluated WavLM and RF across 5 folds and 3 random seeds because they yielded the best performance on mPower large. The results are reported in Table \ref{tab:sex_specific_combined}. For each sex-specific training experiment, we used 50\% of the available data. This corresponds to either the male or female participants while keeping the same testing sets.

\begin{table*}[b!]
\centering
\caption{Comparison of WavLM and Random Forest on female and male subsets from mPower matched and PC-GITA datasets. Values are averaged across folds and seeds in percentage. Asterisks indicate significance levels: * $p<0.05$, ** $p<0.01$, *** $p<0.001$. Grey cells show no significant difference.}
\label{tab:sex_specific_combined}
\begin{tabular}{|c|c|c|c|c|}
\hline
Model & Subset & F1 & F1 female & F1 male \\
\hline
\textbf{WavLM} & male mPower matched & \cellcolor{gray!30}47.99 $\pm$ 3.08 & \cellcolor{gray!30}49.06 $\pm$ 8.04 & \cellcolor{gray!30}46.39 $\pm$ 9.42 \\
\textbf{WavLM} & female mPower matched & \cellcolor{gray!30}54.70 $\pm$ 5.96 & \cellcolor{gray!30}55.99 $\pm$ 8.47 & \cellcolor{gray!30}53.23 $\pm$ 7.08 \\
\textbf{WavLM} & male PC-GITA & \cellcolor{gray!30}40.03 $\pm$ 1.79 & \cellcolor{gray!30}42.57 $\pm$ 3.58 & \cellcolor{gray!30}37.50 $\pm$ 0.00 \\
\textbf{WavLM} & female PC-GITA & \cellcolor{gray!30}62.80 $\pm$ 1.98 & \cellcolor{gray!30}65.90 $\pm$ 5.37 & \cellcolor{gray!30}58.30 $\pm$ 0.00 \\
\hline
\textbf{Random Forest} & male mPower matched & \cellcolor{gray!30}48.19 $\pm$ 4.31 & \cellcolor{gray!30}49.15 $\pm$ 2.59 & \cellcolor{gray!30}45.36 $\pm$ 7.75 \\
\textbf{Random Forest} & female mPower matched & \cellcolor{gray!30}45.26 $\pm$ 4.04 & \cellcolor{gray!30}47.07 $\pm$ 7.17 & \cellcolor{gray!30}42.49 $\pm$ 4.19 \\
\textbf{Random Forest} & male PC-GITA & \cellcolor{gray!30}45.81 $\pm$ 3.75 & \cellcolor{gray!30}23.08 $\pm$ 0.00 & \cellcolor{gray!30}65.60 $\pm$ 7.25 \\
\textbf{Random Forest} & female PC-GITA & \cellcolor{gray!30}52.91 $\pm$ 6.16 & \cellcolor{gray!30}53.81 $\pm$ 11.43 & \cellcolor{gray!30}42.22 $\pm$ 9.50 \\
\hline
\end{tabular}
\end{table*}

From Table \ref{tab:sex_specific_combined}, we notice that, overall, the performance on sex-specific training datasets is lower compared to figure \ref{fig:datasets_compare} when training was done on both sexes. However, none of the sex-specific training configurations produced statistically significant results for either sex, dataset, or model. These results suggest that restricting the training data to a single sex does not improve generalization to both sexes and may reduce the diversity of speech characteristics available for learning robust representations. However, the observed differences should be interpreted cautiously because none of the results reached statistical significance  Appendix ~\ref{app:results} tables \ref{tab:app5},\ref{tab:app6}.

Speech impairment in PD is highly heterogeneous across patients, varying in type, severity, and progression \cite{ho2012speech}. This variability can be amplified by sex differences, hormonal influences on dopaminergic systems, and anatomical variations in vocal tract structure \cite{miller2010gender,titze1989physiologic}.

\section{Conclusion}
In this study, we investigated the sex differences when detecting Parkinson's
Disease from speech using Deep Learning and traditional Machine Learning models.
We concluded that in the larger datasets, classification performance was
significantly above chance for female participants, whereas performance for male
participants was not. The consistency of this pattern across different models
and feature representations suggests that the observed differences are not
specific to a single algorithm and may instead be related to dataset
characteristics and sex-related differences in the manifestation of PD speech
\cite{mehrabi2021survey,larrazabal2020gender}. However, the smaller datasets did
not produce results significantly above chance.

In addition, training models on data from only one sex and evaluating them on
both sexes did not reliably generalize across the evaluated models and datasets.
This highlights the risks of deploying models trained on demographically
restricted datasets without evaluating their performance across demographic
groups and testing the statistical significance of those differences.

Although we explicitly considered fairness by using datasets balanced by sex and
disease status and by applying cross-validation to maintain consistent
representation across folds, sex-related performance differences persisted.
These results suggest that demographic balance alone may not be sufficient to
ensure fair model performance. Other dataset characteristics, including
variability in symptom presentation, recording conditions, disease severity, and
clinical characteristics, should also be considered when evaluating algorithmic
fairness.

Future work should include experimenting with other speech tasks, languages, recording settings, disease severity to identify if the sex differences are general or specific to certain conditions.

\appendix
\section{Additional Metrics}
\label{app:results}

The following tables report additional metrics computed alongside the F1 scores presented in the main text.
\onecolumn 
\begin{table*}[b!] 
\centering 
\begin{tabular}{|c|c|c|c|c|c|c|} 
\hline Models & Accuracy & F1 & F1 male & F1 female \\ 
\hline 
WavLM & 69.04 $\pm$ 5.86 & \cellcolor{gray!30}68.61 $\pm$ 6.61  & \cellcolor{gray!30}53.65 $\pm$ 8.39 & 83.12 $\pm$ 9.36 \textsuperscript{***} \\ 
FBanks & 63.52 $\pm$ 5.81 & \cellcolor{gray!30}63.36 $\pm$ 5.87  & \cellcolor{gray!30}43.30 $\pm$ 7.73 & 82.63 $\pm$ 10.57 \textsuperscript{***}\\ 
Wav2Vec2.0 & 57.2 $\pm$ 7.28 & \cellcolor{gray!30}55.66 $\pm$ 8.40 & \cellcolor{gray!30}49.30 $\pm$ 7.39 & 60.79 $\pm$ 13.05 \textsuperscript{***} \\ 
Whisper & 67.76 $\pm$ 5.95 & \cellcolor{gray!30}66.54 $\pm$ 6.38 & \cellcolor{gray!30}56.00 $\pm$ 7.40 & 76.38 $\pm$ 9.36 \textsuperscript{***}\\ 
Random Forest & 66.64 $\pm$ 3.36 & \cellcolor{gray!30}66.61 $\pm$ 3.36 & \cellcolor{gray!30}46.80 $\pm$ 5.83 & 85.02 $\pm$ 3.79 \textsuperscript{***}\\ 
\hline 
\end{tabular} 
\caption{Performance of 5 feature extractors on mPower large. Asterix indicates significance levels: * p < 0.05, ** p < 0.01, *** p < 0.001. Grey cells
show no significant difference.} 
\label{tab:accuracy}
\end{table*}

\begin{table*}[b!]
\centering
\begin{tabular}{|c|c|c|c|c|c|c|}
\hline
Datasets & Accuracy & F1 & F1 male & F1 female  \\
\hline
mPower matched & 56.87 $\pm$ 1.39 & \cellcolor{gray!30}56.54$\pm$1.47  & \cellcolor{gray!30}51.71$\pm$3.93  & 61.16$\pm$3.39 \textsuperscript{***} \\
mPower small & 53.00 $\pm$ 7.02 & \cellcolor{gray!30}50.01 $\pm$8.99 & \cellcolor{gray!30}48.81$\pm$12.66  & \cellcolor{gray!30}49.45$\pm$16.25   \\
mPower large & 69.04 $\pm$ 5.86 & \cellcolor{gray!30}68.61 $\pm$ 6.61 & \cellcolor{gray!30}53.65 $\pm$ 8.39 & 83.12 $\pm$ 9.36 \textsuperscript{***} \\
PC-GITA  &53.33 $\pm$6.99 & \cellcolor{gray!30}51.40$\pm$6.69 & \cellcolor{gray!30}50.57$\pm$12.58  & \cellcolor{gray!30}51.18$\pm$7.95 \\
\hline
\end{tabular}
\caption{Performance of WavLM on mPower (matched, small, and large) and PC-GITA. 
Asterix indicates significance levels: * $p<0.05$, ** $p<0.01$, *** $p<0.001$. Grey cells show no significant difference. }
\label{tab:datasets_dif}
\end{table*}

\begin{table*}[b!]
\centering
\small
\begin{tabular}{|l|c|c|c|c|}
\hline
\textbf{Subset} & \textbf{Accuracy} & \textbf{F1} & \textbf{Precision} & \textbf{Recall} \\
\hline
male mPower matched & 49.47 $\pm$ 3.04 & \cellcolor{gray!30}47.99 $\pm$ 3.08 & 49.49 $\pm$ 3.27 & 49.47 $\pm$ 3.04 \\
female mPower matched & 55.40 $\pm$ 5.83 & \cellcolor{gray!30}54.70 $\pm$ 5.96 & 55.72 $\pm$ 5.99 & 55.40 $\pm$ 5.83 \\
male PC-GITA & 48.85 $\pm$ 3.63 & \cellcolor{gray!30}38.54 $\pm$ 6.86 & 35.11 $\pm$ 12.14 & 48.85 $\pm$ 3.63 \\
female PC-GITA & 62.33 $\pm$ 7.29 & \cellcolor{gray!30}60.97 $\pm$ 7.70 & 64.79 $\pm$ 8.84 & 62.33 $\pm$ 7.29 \\
\hline
\multicolumn{5}{l}{\textit{(a) Overall metrics}}
\end{tabular}
\end{table*}

\begin{table*}[b!]
\centering
\scriptsize
\setlength{\tabcolsep}{4pt}
\begin{tabular}{|l|c|c|c|c|c|c|}
\hline
\textbf{Subset} & \multicolumn{3}{c|}{\textbf{Female}} & \multicolumn{3}{c|}{\textbf{Male}} \\
\cline{2-7}
 & \textbf{F1} & \textbf{Prec.} & \textbf{Rec.} & \textbf{F1} & \textbf{Prec.} & \textbf{Rec.} \\
\hline
male mPower matched & \cellcolor{gray!30}49.06 $\pm$ 8.04 & 51.50 $\pm$ 9.04 & 51.06 $\pm$ 7.96 & \cellcolor{gray!30}46.39 $\pm$ 9.42 & 47.15 $\pm$ 9.64 & 47.90 $\pm$ 8.51 \\
female mPower matched & \cellcolor{gray!30}55.99 $\pm$ 8.47 & 56.66 $\pm$ 8.70 & 56.38 $\pm$ 8.50 & \cellcolor{gray!30}53.23 $\pm$ 7.08 & 54.93 $\pm$ 7.67 & 54.40 $\pm$ 6.80 \\
male PC-GITA & \cellcolor{gray!30}38.50 $\pm$ 6.36 & 35.62 $\pm$ 12.35 & 49.23 $\pm$ 2.77 & \cellcolor{gray!30}38.35 $\pm$ 7.90 & 34.71 $\pm$ 12.59 & 48.46 $\pm$ 5.55 \\
female PC-GITA & \cellcolor{gray!30}66.89 $\pm$ 7.64 & 70.97 $\pm$ 7.98 & 68.00 $\pm$ 6.76 & \cellcolor{gray!30}54.27 $\pm$ 11.58 & 58.30 $\pm$ 15.68 & 56.67 $\pm$ 10.47 \\
\hline
\multicolumn{7}{l}{\textit{(b) Gender-specific metrics}}
\end{tabular}
\caption{WavLM on female and male subsets from mPower matched and PC-GITA in percent averaged across 5 folds and 3 seeds. Grey cells
show no significant difference.}
\label{tab:app5}
\end{table*}

\begin{table*}[b!]
\centering
\small
\begin{tabular}{|l|c|c|c|c|}
\hline
\textbf{Subset} & \textbf{Accuracy} & \textbf{F1} & \textbf{Precision} & \textbf{Recall} \\
\hline
male mPower matched & 49.68 $\pm$ 4.45 & \cellcolor{gray!30}48.19 $\pm$ 4.31 & 49.80 $\pm$ 4.89 & 49.68 $\pm$ 4.45 \\
female mPower matched & 45.49 $\pm$ 3.92 & \cellcolor{gray!30}45.26 $\pm$ 4.04 & 45.39 $\pm$ 4.03 & 45.49 $\pm$ 3.92 \\
male PC-GITA & 48.00 $\pm$ 3.56 & \cellcolor{gray!30}45.81 $\pm$ 3.75 & 47.60 $\pm$ 4.26 & 48.00 $\pm$ 3.56 \\
female PC-GITA & 54.67 $\pm$ 5.91 & \cellcolor{gray!30}52.91 $\pm$ 6.16 & 55.51 $\pm$ 6.78 & 54.67 $\pm$ 5.91 \\
\hline
\multicolumn{5}{l}{\textit{(a) Overall metrics}}
\end{tabular}
\end{table*}

\begin{table*}[b!]
\centering
\scriptsize
\setlength{\tabcolsep}{4pt}
\begin{tabular}{|l|c|c|c|c|c|c|}
\hline
\textbf{Subset} & \multicolumn{3}{c|}{\textbf{Female}} & \multicolumn{3}{c|}{\textbf{Male}} \\
\cline{2-7}
 & \textbf{F1} & \textbf{Prec.} & \textbf{Rec.} & \textbf{F1} & \textbf{Prec.} & \textbf{Rec.} \\
\hline
male mPower matched & \cellcolor{gray!30}45.36 $\pm$ 7.75 & 50.80 $\pm$ 11.01 & 49.79 $\pm$ 7.90 & \cellcolor{gray!30}49.15 $\pm$ 2.59 & 49.60 $\pm$ 2.82 & 49.58 $\pm$ 2.69 \\
female mPower matched & \cellcolor{gray!30}42.49 $\pm$ 4.19 & 42.76 $\pm$ 4.19 & 43.14 $\pm$ 3.97 & \cellcolor{gray!30}47.07 $\pm$ 7.17 & 48.15 $\pm$ 8.93 & 47.84 $\pm$ 7.48 \\
male PC-GITA & \cellcolor{gray!30}23.08 $\pm$ 0.00 & 18.75 $\pm$ 0.00 & 30.00 $\pm$ 0.00 & \cellcolor{gray!30}65.60 $\pm$ 7.25 & 66.74 $\pm$ 7.36 & 66.00 $\pm$ 7.12 \\
female PC-GITA & \cellcolor{gray!30}53.81 $\pm$ 11.43 & 54.85 $\pm$ 12.44 & 54.67 $\pm$ 10.87 & \cellcolor{gray!30}42.22 $\pm$ 9.50 & 49.63 $\pm$ 26.33 & 54.67 $\pm$ 4.99 \\
\hline
\multicolumn{7}{l}{\textit{(b) Gender-specific metrics}}
\end{tabular}
\caption{Random Forest on female and male subsets from mPower matched and PC-GITA in percent. Grey cells show no significant difference.}
\label{tab:app6}
\end{table*}

\twocolumn            

\section*{Acknowledgment}
We acknowledge support from the Digital Research Alliance of Canada (alliancecan.ca). These data were contributed by users of the Parkinson mPower mobile application as part of the mPower study developed by Sage Bionetworks and described in Synapse [doi:10.7303/syn4993293].

\end{document}